\documentclass{article}
\usepackage{spconf,amsmath,epsfig}
\ninept

\title{Blind Interference Suppression and Power Adjustment with Alternating Optimization
for Cooperative DS-CDMA Networks
 \vspace{-0.5em}} \name{Rodrigo C. de Lamare
\vspace{-0.85em}}
\address{Communications Research Group, Department of
Electronics, \\
    University of York, United Kingdom \\
    Email: rcdl500@ohm.york.ac.uk \vspace{-0.85em}
    }
\begin{document}
\maketitle

\begin{abstract} \vspace{-0.25em}
This work presents blind joint interference suppression and power
allocation algorithms for DS-CDMA networks with multiple relays
and decode and forward protocols. A scheme for joint allocation of
power levels across the relays subject to group-based power
constraints and the design of linear receivers for interference
suppression is proposed. A code-constrained constant modulus (CCM)
design for the receive filters and the power allocation vectors is
devised along with a blind channel estimator. In order to solve
the proposed optimization efficiently, an alternating optimization
strategy is presented with recursive least squares (RLS)-type
algorithms for estimating the parameters of the receiver, the
power allocation and the channels. Simulations show that the
proposed algorithms obtain significant gains in capacity and
performance over existing schemes. \vspace{-0.95em}
\end{abstract}
\begin{keywords}
DS-CDMA, cooperative systems, optimization methods, blind
algorithms, resource allocation.
\end{keywords}\vspace{-0.85em}

\section{Introduction}
Multi-antenna wireless communication systems can exploit the
spatial diversity in wireless channels, mitigating the effects of
fading and enhancing their performance. Due to size and cost of
mobile terminals, it is usually impractical to equip them with
multiple antennas. However, spatial diversity gains can be
obtained when single-antenna terminals establish a distributed
antenna array via cooperation \cite{sendonaris}-\cite{laneman04}.
In a cooperative system, terminals or users relay signals to each
other in order to propagate redundant copies of the same signals
to the destination user or terminal. To this end, the designer
must use a cooperation protocol such as amplify-and-forward (AF)
\cite{laneman04} and decode-and-forward (DF)
\cite{laneman04,huang}.

The use of cooperative diversity and multiple hops is key for
networks that need to increase the link reliability and extend their
coverage \cite{laneman04}. Prior work on cooperative multiuser
DS-CDMA networks has focused on the assessment of the impact of
multiple access interference (MAI) and intersymbol interference
(ISI), the problem of partner selection \cite{huang,venturino}, the
bit error rate (BER) and outage performance analyses \cite{vardhe},
resource allocation \cite{luo,long} and training-based joint power
allocation and interference mitigation strategies
\cite{delamare_jpais,joung}. However, these strategies require a
significant amount of training data and signalling, decreasing
substantially the spectral efficiency of cooperative networks. This
problem is central to ad-hoc and sensor networks \cite{souryal} that
employ spread spectrum systems and multiple hops. This calls for
methods to decrease the amount of signalling and training in
cooperative wireless networks.

In this work, blind joint interference suppression and power
allocation algorithms for DS-CDMA networks with multiple relays and
DF protocols are proposed. A blind scheme that jointly considers the
power allocation across the relays subject to group-based power
constraints and the design of linear receivers for interference
suppression is proposed. The idea of a group-based power allocation
constraint is shown to yield close to optimal performance, while
keeping the signalling and complexity requirements low. A
code-constrained constant modulus (CCM) design
\cite{delamare_ccm}-\cite{barc} for the receive filters and the
power allocation vectors is developed along with a blind channel
estimator. The CCM design is adopted as it achieves a performance
close to training-based algorithms. In order to solve the proposed
optimization problem efficiently, an alternating optimization
strategy is presented with recursive least squares (RLS)-type
algorithms for estimating the parameters of the receiver, the power
allocation and the channels.

The paper is organized as follows. Section 2 describes a cooperative
DS-CDMA system model with multiple relays. Section 3 formulates the
problem, the CCM design of the receive filters and the power
allocation vectors subject to a group-based power allocation
constraint, and a blind channel estimator. Section 4 presents the
alternating optimization strategy along with RLS-type algorithms for
estimating the parameters of the receiver, the power allocation and
the channels. Section 5 presents and discusses the simulation
results and Section 6 draws the conclusions of this
work.\vspace{-0.45em}

\section{Cooperative DS-CDMA Network Model}

Consider a synchronous DS-CDMA network with multipath channels, QPSK
modulation, $K$ users, $N$ chips per symbol and $L$ as the maximum
number of propagation paths for each link. The network is equipped
with a DF protocol that allows communication in multiple hops using
$n_r$ fixed relays in a repetitive fashion. We assume that the
source node or terminal transmits data organized in packets with $P$
symbols, the system can coordinate cooperative transmissions, and
the linear receivers at the relay and destination terminals are
synchronized with their desired signals. The received signals are
filtered by a matched filter, sampled at chip rate and organized
into $M \times 1$ vectors ${\boldsymbol r}_{sd}[m_1]$, ${\boldsymbol
r}_{sr_i}[m_1]$ and ${\boldsymbol r}_{r_id}[m_j]$, which describe
the signal received from the source to the destination, the source
to the relays, and the relays to the destination, respectively,
\vspace{-0.25em}
\begin{equation}
\begin{split}
{\boldsymbol r}_{sd}[m_1] & = \sum_{k=1}^K  a_{sd}^k[m_1]
{\boldsymbol C}_k {\boldsymbol h}_{sd,k}[m_1]b_k[m_1]  \\ & \quad +
{\boldsymbol \eta}_{sd}[m_1] + {\boldsymbol n}_{sd}[m_1],
\\
{\boldsymbol r}_{sr_j}[m_1] & = \sum_{k=1}^K a_{sr_j}^k[m_1]
{\boldsymbol C}_k {\boldsymbol h}_{sr_j,k}[m_1] {b}_k[m_1] \\ &
\quad + {\boldsymbol \eta}_{sr_j}[m_1] + {\boldsymbol
n}_{sr_j}[m_1],
\\
{\boldsymbol r}_{r_jd}[m_j] & = \sum_{k=1}^K a_{r_jd}^k[m_j]
{\boldsymbol C}_k {\boldsymbol h}_{r_jd,k}[m_j] \tilde{b}_k[m_j] \\
& \quad + {\boldsymbol \eta}_{r_jd}[m_j] + {\boldsymbol
n}_{r_jd}[m_j], \label{rvec}
\end{split}
\end{equation}
where $M=N+L-1$, $m_j = (j-1)P+1, \ldots, j P$, $i =1, \ldots, P$,
$j = 1 ~ \ldots, n_p$, $P$ is the number of symbols in the packet,
$n_p=n_r+1$ is the number of transmission phases or hops, $n_r$ is
the number of relays, $m_j$ is the index of original and relayed
signals, ${\boldsymbol n}_{sd}[m_1]$, ${\boldsymbol n}_{sr_j}[m_1]$
and ${\boldsymbol n}_{r_jd}[m_j]$ are zero mean complex Gaussian
vectors with variance $\sigma^2$ generated at the receivers of the
destination and the relays from different links, and the vectors
${\boldsymbol \eta}_{sd}[m_1]$, ${\boldsymbol \eta}_{sr_j}[m_1]$ and
${\boldsymbol \eta}_{r_jd}[m_j]$ represent the intersymbol
interference (ISI). The quantities ${b}_k[m_1]$ and
$\tilde{b}_k[m_j]$ represent the original and reconstructed symbols
by the DF protocol at the relays, respectively. The amplitudes of
the source to destination, source to relay and relay to destination
links for user $k$ are denoted by $a_{sd}^k[m_1]$, $a_{sr_j}^k[m_1]$
and $a_{r_jd}^k[m_j]$, respectively. The $M \times L$ matrix
${\boldsymbol C}_k$ contains versions of the signature sequences of
each user shifted down by one position at each column as described
by
\begin{equation}
{\boldsymbol C}_k = \left[\begin{array}{c c c }
c_{k}(1) &  & {\bf 0} \\
\vdots & \ddots & c_{k}(1)  \\
c_{k}(N) &  & \vdots \\
{\bf 0} & \ddots & c_{k}(N)  \\
 \end{array}\right],
\end{equation}
where ${\boldsymbol c}_k = \big[c_{k}(1), ~c_{k}(2),~ \ldots,~
c_{k}(N) \big]$ stands for the signature sequence of user $k$, the
$L \times 1$ channel vectors  from source to destination, source to
relay, and relay to destination are ${\boldsymbol h}_{sd,k}[m_1]$,
${\boldsymbol h}_{sr_j,k}[m_1]$, ${\boldsymbol h}_{r_jd,k}[m_j]$,
respectively. By collecting the data vectors in (\ref{rvec})
(including the links from relays to the destination) into a
$(n_r+1)M \times 1$ received vector at the destination we obtain {
\begin{equation}
\begin{split}
 \hspace{-0.35em}\left[\hspace{-0.25em}\begin{array}{c}
 \hspace{-0.25em} {\boldsymbol r}_{sd}[m_1] \\
  \hspace{-0.25em} {\boldsymbol r}_{r_{1}d}[m_2] \\
  \hspace{-0.25em} \vdots \\
  \hspace{-0.25em} {\boldsymbol r}_{r_{n_r}d}[m_{n_p}]
\end{array} \hspace{-0.25em}\right] & = \left[\hspace{-0.35em}\begin{array}{l}
 \hspace{-0.1em} \sum_{k=1}^K  a_{sd}^k[m_1] {\boldsymbol C}_k {\boldsymbol h}_{sd,k}[m_1]b_k[m_1] \\
  \hspace{-0.1em}\sum_{k=1}^K  a_{{r_1}d}^k[m_2] {\boldsymbol C}_k {\boldsymbol h}_{{r_1}d,k}[m_2]{\tilde b}_k^{{r_1}d}[m_2] \\
  \vdots \\
  \hspace{-0.1em}\sum_{k=1}^K  a_{{r_{n_r}}d}^k[m_{n_p}] {\boldsymbol C}_k {\boldsymbol h}_{r_{n_r}d,k}[m_{n_p}]{\tilde b}_k^{{r_{n_r}}d}[m_{n_p}]
\end{array} \hspace{-0.5em} \right] \\ & \quad + {\boldsymbol \eta}[i] + {\boldsymbol n}[i]
\end{split}
\end{equation}} \vspace{-0.75em}
Rewriting the above signals in a compact form yields
\begin{equation}
\begin{split}
{\boldsymbol r}[i] & 
= \sum_{k=1}^{K}  \widetilde{\boldsymbol B}_k[i]
\widetilde{\boldsymbol A}_k[i] \widetilde{\boldsymbol {\mathcal
C}}_k {\boldsymbol h}_k[i]+ {\boldsymbol \eta}[i] + {\boldsymbol
n}[i]
\\ & = \sum_{k=1}^{K}  {\boldsymbol P}_k[i]
{\boldsymbol B}_k[i] {\boldsymbol a}_k[i]+ {\boldsymbol \eta}[i] +
{\boldsymbol n}[i] 
, \label{recdata}
\end{split}
\end{equation}
where the $(n_r+1)M \times (n_r+1)L$ matrix
$\widetilde{\boldsymbol {\mathcal C}}_k = {\rm diag} \{
{\boldsymbol C}_k \ldots {\boldsymbol C}_k \}$ contains copies of
${\boldsymbol C}_k$ shifted down by $M$ positions for each group of $L$ columns and zeros elsewhere. 
The $(n_r+1)L \times 1$ vector ${\boldsymbol h}_k[i]$ contains the
channel gains of the links between the source, the relays and the
destination, and ${\boldsymbol p}_k[i]=\widetilde{\boldsymbol
{\mathcal C}}_k {\boldsymbol h}_k[i]$ is the effective signature for
user $k$. The $(n_r+1) \times (n_r+1)$ diagonal matrix ${\boldsymbol
B}_k[i] = {\rm diag}(b_k[m_1]~ {\tilde b}_k^{{r_1}d}[m_2] \ldots
{\tilde b}_k^{{r_{n_r}}d}[m_{n_p}]) $ contains the symbols
transmitted from the source to the destination ($b_k[i]$) and the
$n_r$ symbols transmitted from the relays to the destination
(${\tilde b}_k^{{r_1}d}[m_2] \ldots {\tilde
b}_k^{{r_{n_r}}d}[m_{n_r}]$) on the main diagonal, and the $(n_r+1)M
\times (n_r+1)M$ diagonal matrix $\widetilde{\boldsymbol B}_k[i] =
{\rm diag}(b_k[m_1]\bigotimes {\boldsymbol I}_M~ {\tilde
b}_k^{{r_1}d}[m_2]\bigotimes {\boldsymbol I}_M \ldots {\tilde
b}_k^{{r_{n_r}}d}[m_{n_p}]\bigotimes {\boldsymbol I}_M)$, where
$\bigotimes$ denotes the Kronecker product and ${\boldsymbol I}_M$
is an identity matrix with dimension $M$. The $(n_r+1) \times 1$
power allocation vector ${\boldsymbol
a}_k[i]=[a_{sd}^k[m_1]~a_{{r_1}d}^k[m_2]\ldots
a_{{r_{n_r}}d}^k[m_{n_p}]]^T$ has the amplitudes of the links, the
$(n_r+1) \times (n_r+1)$ diagonal matrix ${\boldsymbol A}_k[i]$ is
given by ${\boldsymbol A}_k[i] = {\rm diag} \{ {\boldsymbol
a}_k[m_1] \}$, and the $(n_r+1)M \times (n_r+1)M$ diagonal matrix
$\widetilde{\boldsymbol A}_k[i]= [a_{sd}^k[m_1]\bigotimes
{\boldsymbol I}_M~a_{{r_1}d}^k[m_2]\bigotimes {\boldsymbol I}_M
\ldots a_{{r_{n_r}}d}^k[m_{n_p}]\bigotimes {\boldsymbol I}_M]^T $.
The $(n_r+1)M \times (n_r+1)$ matrix ${\boldsymbol P}_k$ has copies
of the effective signature ${\boldsymbol p}_k[i]$ shifted down by
$M$ positions for each column and zeros elsewhere. The $(n_r+1)M
\times 1$ vector ${\boldsymbol \eta}[i]$ represents the ISI terms
and the $(n_r+1)M \times 1$ vector ${\boldsymbol n}[i]$ has the
noise components.

\section{Proposed Blind Receiver Design, Power Allocation and Channel Estimation}

In this section, a joint blind receiver design and power
allocation strategy is proposed using the CCM approach and
group-based power constraints along with a blind channel
estimator. To this end, the $(n_r+1)M \times 1$ received vector in
(\ref{recdata}) can be expressed as
\begin{equation}
{\boldsymbol r}[i] = {\boldsymbol P}_{\mathbf {\mathcal S}}[i]
{\boldsymbol B}_{\mathbf {\mathcal S}}[i] {\boldsymbol
a}_{{\mathbf {\mathcal S}},k}[i] + \sum_{k \neq {\mathbf {\mathcal
S}}} {\boldsymbol P}_k[i] {\boldsymbol B}_k[i] {\boldsymbol
a}_k[i]+ {\boldsymbol \eta}[i] + {\boldsymbol n}[i],
\label{recdatag}
\end{equation}
where ${\mathbf {\mathcal S}} = \{{\mathcal S}_1, {\mathcal S}_2,
\ldots, {\mathcal S}_G \}$ denotes the group of $G$ users to
consider in the design. The $(n_r+1)M \times G(n_r+1)$ matrix
${\boldsymbol P}_{\mathbf {\mathcal S}} = [ {\boldsymbol
P}_{{\mathcal S}_1} ~ {\boldsymbol P}_{{\mathcal S}_2} ~ \ldots ~
{\boldsymbol P}_{{\mathcal S}_G} ]$ contains the $G$ effective
signatures of the group of users. The $G(n_r+1) \times G(n_r+1)$
diagonal matrix ${\boldsymbol B}_k[i] = {\rm diag}(b_{{\mathcal
S}_1}[i]~ {\tilde b}_{{\mathcal S}_1}^{{r_1}d}[i] \ldots {\tilde
b}_{{\mathcal S}_1}^{{r_n}d}[i]~ \ldots ~ b_{{\mathcal S}_G}[i]~
{\tilde b}_{{\mathcal S}_G}^{{r_1}d}[i] \ldots {\tilde
b}_{{\mathcal S}_G}^{{r_n}d}[i])$ contains the symbols transmitted
from the sources to the destination and from the relays to the
destination of the $G$ users in the group on the main diagonal,
the $G(n_r+1) \times 1$ power allocation vector ${\boldsymbol
a}_{{\mathbf {\mathcal S}},k}[i]=[a_{sd}^{{\mathcal
S}_1}[i]~a_{{r_1}d}^{{\mathcal S}_1}[i] \ldots
a_{{r_{n_r}}d}^{{\mathcal S}_1}[i],~ \ldots, ~ a_{sd}^{{\mathcal
S}_G}[i]~a_{{r_1}d}^{{\mathcal S}_G}[i] \ldots
a_{{r_{n_r}}d}^{{\mathcal S}_G}[i]]^T$ of the amplitudes of the
links used by the $G$ users in the group.

\subsection{Blind CCM Receiver Design and Power Allocation Scheme with
Group-Based Constraints}

The linear interference suppression for user $k$ is performed by
the receive filter ${\boldsymbol w}_k[i]=[ {w}_{k,1}[i],~ \ldots,
~ {w}_{k,(n_r+1)M}[i]]$ with $(n_r+1)M$ coefficients on the
received data vector ${\boldsymbol r}[i]$ and yields
\begin{equation}
z_k[i] = {\boldsymbol w}_k^H[i] {\boldsymbol r}[i],
\end{equation}
where $z_k[i]$ is an estimate of the symbols, which are processed
by a slicer $Q(\cdot)$ that performs detection and obtains
$\hat{b}_k[i] = Q (z_k[i])$.

Let us now detail the CCM-based design of the receivers for user
$k$ represented by ${\boldsymbol w}_k[i]$ and for the computation
of the $G(n_r +1) \times 1$ power allocation vector ${\boldsymbol
a}_{{\mathbf {\mathcal S}},k}[i]$. This problem can be cast as
\begin{equation}
\begin{split}
\hspace{-0.5em}[ {\boldsymbol w}_{k}^{\rm opt}, ~{\boldsymbol a}_{{{\mathbf
{\mathcal S}},k}}^{\rm opt}  ] & = \arg \min_{{\boldsymbol
w}_k[i], {\boldsymbol a}_{{\mathbf {\mathcal S}},k}[i]} ~
E[ (| {\boldsymbol w}^H_k[i]{\boldsymbol r}[i] |^2-1)^2 ] \\
\hspace{-0.5em} {\rm subject ~to~} & {\boldsymbol a}_{{\mathbf {\mathcal
S}},k}^H[i] {\boldsymbol a}_{{\mathbf {\mathcal S}},k}[i] = P_{G}
~ {\rm and}~ {\boldsymbol w}^H_k[i]{\boldsymbol p}_k[i]= \nu,
\label{probg}
\end{split}
\end{equation}
where $\nu$ is a parameter used to enforce convexity
\cite{delamare_mimoccm}. The CCM expressions for the receive
filter ${\boldsymbol w}_{k}[i]$ and the power allocation vector
${\boldsymbol a}_{{\mathbf {\mathcal S}},k}[i]$ can be obtained
with the method of Lagrange multipliers which transforms (\ref{probg})
into the Lagrangian function {\small
\begin{equation}
\begin{split}
{\mathcal L}_k & = E\big[ \Big( |{\boldsymbol w}_k^H[i] \big(
{\boldsymbol P}_{\mathbf {\mathcal S}}[i] {\boldsymbol B}_{\mathbf
{\mathcal S}}[i] {\boldsymbol a}_{{\mathbf {\mathcal S}},k}[i] \\
& \quad  + \sum_{k \neq {\mathbf {\mathcal S}}} {\boldsymbol
P}_k[i] {\boldsymbol B}_k[i] {\boldsymbol a}_k[i]   + {\boldsymbol
\eta}[i] + {\boldsymbol n}[i]\big) |^2 -1 \Big)^2 \big]   \\
& \quad + \lambda_k ({\boldsymbol a}_{{\mathbf {\mathcal
S}},k}^H[i]{\boldsymbol a}_{{\mathbf {\mathcal S}},k}[i] - P_{G})  +
\rho_k( {\boldsymbol w}^H_k[i]{\boldsymbol p}_k[i]- \nu) ,
\label{lagt}
\end{split}
\end{equation}}
where $\lambda_k$ and $\rho_k$ are Lagrange multipliers. An
expression for ${\boldsymbol a}_{{\mathbf {\mathcal S}},k}[i]$ is
obtained by fixing ${\boldsymbol w}_k[i]$, taking the gradient
terms of the Lagrangian and equating them to zero which yields
\begin{equation}
{\boldsymbol a}_{{\mathbf {\mathcal S}},k}[i] = ( {\boldsymbol
R}_{{\mathbf {\mathcal S}},k}[i] + \lambda_k {\boldsymbol I})^{-1}
{\boldsymbol d}_{{\mathbf {\mathcal S}},k}[i] \label{avect}
\end{equation}
where ${\boldsymbol R}_{{\mathbf {\mathcal S}},k}[i] = E[|z_k[i]|^2
{\boldsymbol B}_{\mathbf {\mathcal S}}^H[i]{\boldsymbol
P}_{\mathbf {\mathcal S}}^H[i]  {\boldsymbol w}_k[i] {\boldsymbol
w}^H_k[i]{\boldsymbol P}_{\mathbf {\mathcal S}}[i] {\boldsymbol
B}_{\mathbf {\mathcal S}}[i]]$ is a $G(n_r+1) \times G(n_r+1)$ correlation matrix
and the $G(n_r+1) \times 1$  vector ${\boldsymbol
d}_{{\mathbf {\mathcal S}},k}[i] = E[z_k[i] {\boldsymbol
B}_{\mathbf {\mathcal S}}^H[i]{\boldsymbol P}_{\mathbf {\mathcal
S}}^H[i]  {\boldsymbol w}_k[i]]$ is a
cross-correlation vector. The Lagrange multiplier $\lambda_k$
plays the role of a regularization term and has to be determined
numerically due to the difficulty of evaluating its expression.
Now fixing ${\boldsymbol a}_{{\mathbf {\mathcal S}},k}[i]$, taking
the gradient terms of the Lagrangian and equating them to zero
leads to
\begin{equation}
{\boldsymbol w}_k[i] = {\boldsymbol R}^{-1}_k[i] ({\boldsymbol
d}_k[i] - {\boldsymbol p}_k[i]\gamma_k^{-1}[i] ({\boldsymbol
p}_k^H[i] {\boldsymbol R}^{-1}_k[i]{\boldsymbol d}_k[i] - \nu),
\label{wvect}
\end{equation}
where $\gamma_k[i] = {\boldsymbol p}_k^H[i]{\boldsymbol
R}^{-1}_k[i] {\boldsymbol p}_k[i]$, the correlation matrix is
given by ${\boldsymbol R}_k[i] = E[|z_k[i]|^2{\boldsymbol
r}[i]{\boldsymbol r}^H[i]]$ and ${\boldsymbol d}_k[i] = E[z_k[i]
{\boldsymbol r}[i]] $ is a $(n_r+1)M \times 1$ cross-correlation
vector. The quantities ${\boldsymbol R}_k[i]$ and ${\boldsymbol
d}_k[i]$ depend on the power allocation vector ${\boldsymbol
a}_{{\mathbf {\mathcal S}},k}[i]$. The expressions in
(\ref{avect}) and (\ref{wvect}) do not have a closed-form solution
as they arise from a higher-order optimization. Moreover, the
expressions also depend on each other and require the estimation
of the channel vector ${\boldsymbol h}_k[i]$. Thus, it is
necessary to iterate (\ref{wvect}) and (\ref{avect}) with initial
values to obtain a solution and to estimate the channel. The
network has to convey the information from the group of users
necessary to compute the group-based power allocation including
the filter ${\boldsymbol w}_k[i]$. The expressions in
(\ref{wvect}) and (\ref{avect}) require matrix inversions with
cubic complexity ( $O(((n_r+1)M)^3)$ and $O((K(n_r+1))^3)$.

\subsection{Blind Cooperative Channel Estimation}

In order to blindly estimate the channel in the cooperative system
under study, let us consider the covariance matrix ${\boldsymbol
R}=E[{\boldsymbol r}[i] {\boldsymbol r}^H[i]]$ and the transmitted
signal ${\boldsymbol x}_{k}[i] = {\boldsymbol A}_k[i]{\boldsymbol
B}_k[i]{\boldsymbol p}_k[i]$. Let us now perform an
eigen-decomposition on ${\boldsymbol R}$
\begin{equation}
\begin{split}
{\boldsymbol R} & = \sum_{k=1}^{K} E[{\boldsymbol
x}_k[i]{\boldsymbol x}_k^H[i]] + E[{\boldsymbol \eta}[i]
{\boldsymbol \eta}^H[i]] + \sigma^2 {\boldsymbol I} \\
& = \big[ {\boldsymbol \phi}_s ~{\boldsymbol \phi}_n \big]
\left[\begin{array}{cc} {{\boldsymbol \Lambda}_s + \sigma^2
{\boldsymbol I}} & {\boldsymbol 0} \\  {\boldsymbol 0} & {\sigma^2
{\boldsymbol I}} \end{array}\right]  \big[ {\boldsymbol \phi}_s
~{\boldsymbol \phi}_n \big]^H,
\end{split}
\end{equation}
where ${\boldsymbol \phi}_s$ and ${\boldsymbol \phi}_n$ are the
signal and noise subspaces, respectively. Since ${\boldsymbol
\phi}_s$ and ${\boldsymbol \phi}_n$ are orthogonal, we have the
condition ${\boldsymbol \phi}_n^H {\boldsymbol x}_{k}[i] =
{\boldsymbol \phi}_n^H {\boldsymbol A}_k[i]{\boldsymbol
B}_k[i]{\boldsymbol p}_k[i] = {\boldsymbol \phi}_n^H {\boldsymbol
A}_k[i]{\boldsymbol B}_k[i]{\boldsymbol C}_k{\boldsymbol
h}_k[i]={\boldsymbol 0}$ and hence
\begin{equation}
\begin{split}
\Gamma  = {\boldsymbol h}_k^H[i] \underbrace{{\boldsymbol C}_k^H
{\boldsymbol B}_k^H[i] {\boldsymbol A}_k^H[i] {\boldsymbol \phi}_n
{\boldsymbol \phi}_n^H {\boldsymbol A}_k[i] {\boldsymbol B}_k[i]
{\boldsymbol C}_k}_{\boldsymbol \Upsilon_k} {\boldsymbol h}_k[i]
\end{split}
\end{equation}
The above relation  allows to blindly estimate the channel
${\boldsymbol h}_k[i]$. To this end, we need to compute the
eigenvector corresponding to the smallest eigenvalue of
${\boldsymbol \Upsilon}_k$. It turns out that we can use the fact
that $\lim_{p \rightarrow \infty} ({\boldsymbol R}/\sigma^2)^{-p}
= {\boldsymbol \phi}_n {\boldsymbol \phi}_n^H$
\cite{delamare_mimoccm} and, in practice, it suffices to use $p=1
~{\rm or}~2$. Therefore, to blindly estimate the channel of user
$k$ in the cooperative system we need to solve the optimization
problem
\begin{equation}
\hat{\boldsymbol h}_k[i] = \arg \min_{{\boldsymbol h}_k[i]}
{\boldsymbol h}_k^H[i] {\boldsymbol \Upsilon_k} {\boldsymbol
h}_k[i], ~~ {\rm subject}~~{\rm to} ~~ ||{\boldsymbol h}_k[i]||=1,
\label{cest}
\end{equation}
In what follows, computationally efficient algorithms based on an
alternating optimization strategy will be detailed.

\section{Proposed Adaptive Algorithms}

In this section, we develop joint adaptive RLS-type algorithms using
an alternating optimization strategy for efficiently estimating the
parameters of the receive filters, the power allocation vectors and
the channels. Note that the proposed algorithms did not have
problems with local minima and converge to the desired solutions.

The first task in the proposed scheme is to build the group of $G$
users that will be used for the power allocation and receive
filter design. A RAKE receiver is employed to obtain $z_k^{\rm
RAKE}[i] = ({\boldsymbol C}_k\hat{\boldsymbol
h}_k[i])^H{\boldsymbol r}[i]=\hat{\boldsymbol
p}_k^H[i]{\boldsymbol r}[i]$ and the group is formed according to
\begin{equation}
{\rm compute} ~~{\rm the}~~G~~{\rm largest}~~|z_k^{\rm
RAKE}[i]|,~~ k=1,2, \ldots, K. \label{group}
\end{equation}
The design of the RAKE and the other tasks require channel
estimation. In order to solve (\ref{cest}) efficiently, a variant
of the power method \cite{delamare_ccm} that uses a simple shift
is adopted
\begin{equation}
\hat{\boldsymbol h}_k[i] = ({\boldsymbol I} - \tau_k[i]
\hat{\boldsymbol \Upsilon}_k[i] ) \hat{\boldsymbol h}_k[i-1],
\label{cestpow}
\end{equation}
where $\tau_k[i] = 1/tr[\hat{\boldsymbol \Upsilon}_k[i]]$ and
$\hat{\boldsymbol h}_k[i] \leftarrow \hat{\boldsymbol h}_k[i]/
||\hat{\boldsymbol h}_k[i]||$ to normalize the channel. The
quantity $\hat{\boldsymbol \Upsilon}_k[i]$ is estimated by
$\hat{\boldsymbol \Upsilon}_k[i] = \alpha \hat{\boldsymbol
\Upsilon}[i-1] + {\boldsymbol C}_k^H \hat{\boldsymbol B}_k^H[i]
\hat{\boldsymbol A}_k^H[i] \hat{\boldsymbol R}^{-p}[i]
\hat{\boldsymbol A}_k[i] \hat{\boldsymbol B}_k[i] {\boldsymbol
C}_k$, where $\alpha$ is a forgetting factor that should be close
to $1$ and $\hat{\boldsymbol R}^{-p}[i]$ is computed with the
matrix inversion lemma. The power allocation and receive filter
design problems outlined in (\ref{probg}) are solved by replacing
the expected values in (\ref{avect}) and (\ref{wvect}) with time
averages, and RLS-type algorithms. The approach for allocating the
power within a group is to drop the constraint, estimate the
quantities of interest and then impose the constraint via a
subsequent normalization. The group-based power allocation is
computed by
\begin{equation}
\hat{\boldsymbol a}_{{\mathbf{\mathcal S}},k}[i] =
\hat{\boldsymbol P}_{{\mathbf{\mathcal S}},k}[i-1]
\hat{\boldsymbol d}_{{\mathbf{\mathcal S}},k}[i], \label{arls}
\end{equation}
where \begin{equation} \hat{\boldsymbol d}_{{\mathbf{\mathcal
S}},k}[i] = \alpha \hat{\boldsymbol d}_{{\mathbf{\mathcal
S}},k}[i]+ z_k[i] {\boldsymbol v}_k[i],
\end{equation}
\begin{equation}
{\boldsymbol k}_{{\mathbf{\mathcal S}},k} = \frac{\alpha^{-1}
\hat{\boldsymbol P}_{{\mathbf{\mathcal S}},k}[i-1] z_k[i]
{\boldsymbol v}_k[i]}{1+\alpha^{-1} {\boldsymbol v}_k^H[i]
\hat{\boldsymbol P}_{{\mathbf{\mathcal S}},k}[i-1] {\boldsymbol
v}_k[i] |z_k[i]|^2 },
\end{equation}
\begin{equation}
\hat{\boldsymbol P}_{{\mathbf{\mathcal S}},k}[i] = \alpha^{-1}
\hat{\boldsymbol P}_{{\mathbf{\mathcal S}},k}[i-1] - \alpha^{-1}
z_k^*[i] {\boldsymbol k}_{{\mathbf{\mathcal S}},k}[i] {\boldsymbol
v}_k^H[i] \hat{\boldsymbol P}_{{\mathbf{\mathcal S}},k}[i-1].
\end{equation}
The normalization $\hat{\boldsymbol a}_{{\mathbf{\mathcal
S}},k}[i] \leftarrow P_G  ~\hat{\boldsymbol a}_{{\mathbf{\mathcal
S}},k}[i]/||\hat{\boldsymbol a}_{{\mathbf{\mathcal S}},k}[i]|| $
is then made to ensure the power constraint. The receive filter is
computed by
\begin{equation}
\hat{\boldsymbol w}_k[i] = {\boldsymbol P}_k[i] (\hat{\boldsymbol
d}_k[i] - \hat{\boldsymbol p}_k[i]\hat{\gamma}_k^{-1}[i]
(\hat{\boldsymbol p}_k^H[i]\hat{\boldsymbol
P}_k[i]\hat{\boldsymbol d}_k[i] - \nu), \label{wrls}
\end{equation}
where $\gamma_k^{-1}[i] = \hat{\boldsymbol
p}_k^H[i]\hat{\boldsymbol P}_k[i] \hat{\boldsymbol p}_k[i]$ and
\begin{equation} \hat{\boldsymbol d}_{k}[i] = \alpha
\hat{\boldsymbol d}_{k}[i]+ z_k[i] {\boldsymbol r}[i],
\end{equation}
\begin{equation}
{\boldsymbol k} = \frac{\alpha^{-1} \hat{\boldsymbol P}_{k}[i-1]
z_k[i] {\boldsymbol r}[i]}{1+\alpha^{-1} {\boldsymbol r}^H[i]
\hat{\boldsymbol P}_{k}[i-1] {\boldsymbol r}[i] |z_k[i]|^2 },
\end{equation}
\begin{equation}
\hat{\boldsymbol P}_{k}[i] = \alpha^{-1} \hat{\boldsymbol
P}_{k}[i-1] - \alpha^{-1} z_k^*[i] {\boldsymbol k}[i] {\boldsymbol
r}^H[i] \hat{\boldsymbol P}_{k}[i-1]. \label{mil2}
\end{equation}
The proposed scheme employs the algorithm in (\ref{group}) to
allocate the users in the group and the channel estimation
approach of (\ref{cestpow}). The alternating optimization strategy
uses the recursions (\ref{arls})-(\ref{mil2}) with $1~{\rm or}~2$
iterations per time instant $i$.

\section{Simulations}

The bit error ratio (BER) performance of the proposed blind joint
power allocation and interference suppression (BJPAIS) scheme and
algorithms with group-based constraints (GBC) is assessed. The
BJPAIS scheme and algorithms are compared with blind schemes
without cooperation (BNCIS) \cite{delamare_ccm} and with
cooperation (CIS) using an equal power allocation across the
relays (the power allocation in the BJPAIS scheme is disabled). A
DS-CDMA network with randomly generated spreading codes and a
processing gain $N=16$ is considered. The fading channels are
generated considering a random power delay profile with gains
taken from a complex Gaussian variable with unit variance and mean
zero, $L=5$ paths spaced by one chip, and are normalized for unit
power. The power constraint parameter $P_{A,k}$ is set for each
user so that one can control the SNR (${\rm SNR} =
P_{A,k}/\sigma^2$) and $P_T= P_G + (K-G) P_{A,k}$, whereas it
follows a log-normal distribution for the users with associated
standard deviation equal to $3$ dB. The DF cooperative protocol is
adopted and all the relays and the destination terminal use linear
CCM receivers.

\begin{figure}[!htb]
\begin{center}
\def\epsfsize#1#2{0.925\columnwidth}
\epsfbox{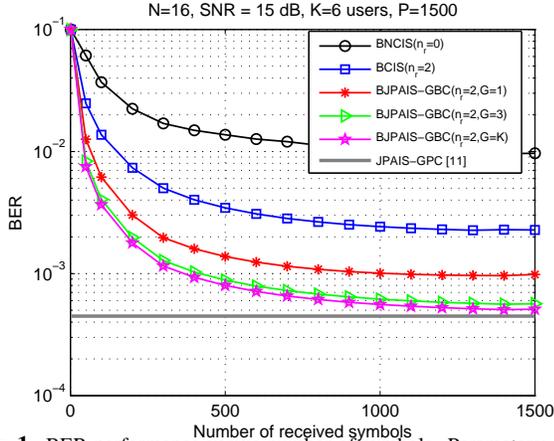} \vspace{-1.75em} \caption{\footnotesize BER
performance versus number of symbols. Parameters: $p=1$,
$\lambda_T=\lambda_k =0.025$ (for MMSE and CCM schemes),
$\alpha=0.998$, $\hat{\boldsymbol P}_{{\mathbf{\mathcal
S}},k}[i]=0.01 {\boldsymbol I}$ and $\hat{\boldsymbol P}_{k}[i]=0.01
{\boldsymbol I}$.} \vspace{-0.85em}\label{fig1}
\end{center}
\end{figure}

The first experiment depicted in Fig. \ref{fig1} shows the BER
performance of the proposed BJPAIS scheme and algorithms against
the BNCIS and BCIS schemes with $n_r=2$ relays. The BJPAIS scheme
is considered with the group-based power constraints (BJPAIS-GBC).
All techniques employ RLS-type algorithms for estimation of the
channels, the receive filters and the power allocation for each
user. The results show that as the group size $G$ is increased the
proposed BJPAIS scheme and algorithms converge to approximately
the same level of the cooperative training-based JPAIS-MMSE scheme
reported in \cite{delamare_jpais}, which employs $G=K$ for power
allocation, and has full knowledge of the channel and the noise
variance.

\begin{figure}[!htb]
\begin{center}
\def\epsfsize#1#2{0.925\columnwidth}
\epsfbox{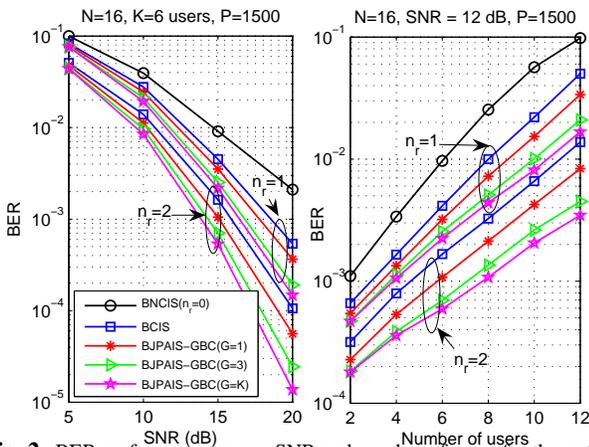} \vspace{-1.75em} \caption{\footnotesize BER
performance versus SNR and number of users for the optimal linear
MMSE detectors. Parameters: $p=1$, $\alpha =0.998$,
$\hat{\boldsymbol P}_{{\mathbf{\mathcal S}},k}[i]=0.01 {\boldsymbol
I}$ and $\hat{\boldsymbol P}_{k}[i]=0.01 {\boldsymbol I}$.}
\vspace{-0.85em}\label{fig2}
\end{center}
\end{figure}

The proposed BJPAIS-GBC scheme is then compared with a
non-cooperative approach (BNCIS) and a cooperative scheme with
equal power allocation (BCIS) across the relays for $n_r=1,2$
relays. The results shown in Fig. \ref{fig2} illustrate the
performance improvement achieved by the BJPAIS scheme and
algorithms, which significantly outperform the BCIS and the BNCIS
techniques. As the number of relays is increased so is the
performance, reflecting the exploitation of the spatial diversity.
In the scenario studied, the proposed BJPAIS-GBC with $G=3$ can
accommodate up to $3$ more users as compared to the BCIS scheme
and double the capacity as compared with the BNCIS for the same
BER performance, without the need for training data. The curves
indicate that the GBC for power allocation with only a few users
is able to attain a performance close to the BJPAIS-GBC with $G=K$
users, while requiring a lower complexity and extra network
signalling. A detailed study of the signalling requirements will
be considered in a future work.

\section{Conclusions}
 \vspace{-0.75em}
This work has proposed the BJPAIS scheme for cooperative DS-CDMA
networks with multiple relays and the DF protocol. A CCM design
for the receive filters and the power allocation with group
constraints has been devised along with a blind channel estimator
and RLS-type algorithms. The results have shown that the BJPAIS
scheme achieves significant gains in performance and capacity over
existing schemes, without requiring training data. Future work
will consider distributed space-time coding and synchronization.

\end{document}